\documentclass{article}
\usepackage{amsmath,amssymb,amsfonts}

\begin{document}

\begin{center}

{\bf FOUR DIMENSIONAL TOPOLOGICAL QUANTUM FIELD THEORY,\\ HOPF CATEGORIES, AND
THE CANONICAL
BASES}

\end{center}

\smallskip
by Louis Crane, Department of Mathematics, KSU

and Igor B. Frenkel, Department of Mathematics, Yale University

\smallskip

{\bf Abstract}: We propose a new combinatorial method of constructing
4D-TQFTs. The method uses a new type of algebraic structure called a
Hopf category. We also outline the construction of a family of Hopf
categories related to the quantum groups, using the canonical bases.

\smallskip

{\bf  INTRODUCTION}

\smallskip

Since the seminal papers of Witten [1], and Atiyah [2], the attention of
many mathematicians and physicists has been drawn to the problem of
constructing topological quantum field theories (TQFTs). The mathematical form
of the definition of a TQFT uses very categorical language, and is strongly
influenced by the definition of a conformal field theory due to Segal [3].The
subject of TQFT
has a dual character, as a new development in algebraic topology on
one hand, and as a branch of quantum field theory on the other. While
physicists have
suggested that various TQFTs exist in various dimensions by
non-rigorous and ill-defined arguments involving path integrals, many
rigorous constructions have been discovered using various
techniques involving abstract algebra [4,5,6,7,8,9]. There has
been a close interaction between the mathematical and physical
approaches, which has led to the rapid development of the subject.

The most exciting theories whose existence has been suggested are the
cousins of Donaldson-Floer (DF) theory [1], which should exist in 4
dimensions, and be sensitive to smooth structure.   So far, neither the
analytic [10]
nor the path integral  nor any algebraic approach has succeeded in
constructing Donaldson-Floer theory, except for special cases.

The only 4D-TQFTs which have been rigorously constructed so far are
Dijkgraaf-Witten theory [11,12,13], which counts representations of $\pi_1$
into a finite group, and Crane-Yetter theory [14], which  can be used to
find the signature, Euler character, second Chern class and cup
product on the second cohomology of the 4 manifold. Both of these
theories, in short, produce classical invariants (although by new
combinatorial formulae), and hence, are not obviously related to
Donaldson-Floer theory.

The purpose of this paper is to propose a new algebraic construction
of a 4D-TQFT, which uses a novel algebraic structure, called a Hopf
category. This would not be very interesting, except that we also
propose a construction of a family of Hopf categories,
intimately related to the canonical bases for the quantum groups. We
conjecture that the  4D-TQFTs   which arise by our new method from our
Hopf categories may be sensitive to smooth structure, and in fact may
be related to DF theory.

As far as the current status of DF theory is concerned, let us make a
hypothetical historical analogy. Suppose somebody had used the
implicit function theorem to show that the dimension of the space of
harmonic forms on a Riemannian manifold was independent of the metric.
One would then have a set of ``analytic'' invariants of closed
manifolds. It would then take some imagining to find a combinatorial
algebraic formula for these invariants from a triangulation.
Of course, once that invariant had been understood as rational
cohomology, nobody would think of de Rham's definition of it as truly
fundamental any more.

DF invariants come from counting instantons, a sort of nonabelian
version of harmonic forms. It would not be too surprising if the
combinatorial version of these invariants utilized noncommutative
algebra.

Let us try to explain some of the ideas which underly our
construction.

The first is that the constructions of particular types of
TQFTs in various dimensions, written in terms of generators and relations, end
up requiring precisely a family of
classical
structures of abstract algebra. Let us give just one simple
example to illustrate this. If we try to construct a 2D-TQFT from
triangulated surfaces by putting labels {i} on edges and attaching
numbers $C_{ijk}$ depending on the edge labels to triangles, then the
topological move:

\smallskip

Figure 1

\smallskip

tells us that we need $C_{ijk}$ to be the structure constants of an
associative algebra (in fact, it tells us a lot more, namely, that we
need a symmetric Frobenius algebra). Schematically, we can draw the associative
law as
follows:

\smallskip

Figure 2

\smallskip

The other topological moves on triangulations
tell us that we actually need an associative algebra with
a rigid structure of a classically familiar type, namely a semisimple
algebra [9].

The second idea motivating our construction is that replacing an
algebraic structure with a similar categorical algebraic structure
lifts the dimension of the corresponding TQFT by 1. For example, we
can construct a 3D-TQFT from a suitable tensor category by analogy
with the construction discussed above. The 3D analog of the move
illustrated above is the famous Stasheff pentagon [15], also known as
the Biedenharn-Eliot relation [16]. This is essentially the
construction in [8].

The third idea leading up to our construction is that it is possible
to use a Hopf algebra to construct a 3D-TQFT, by a process analogous
to the use of an algebra in 2D, combining the use of the algebra
structure on the 2-skeleton with the coproduct structure on the dual
2-skeleton of a 3D triangulation [7].

The combination of the last two ideas might suggest that a categorical
version of a Hopf algebra might pull
us up to 4D-TQFT. We call the new algebraic structures, which can be
viewed as generators and relations for 4D-TQFT, Hopf categories.

To proceed any farther we need a miracle, namely, the existence of
an interesting family of Hopf categories.

The next important input is the existence of the canonical bases [17]
for a special family of Hopf algebras, namely the quantum groups.
These bases are actually an indication of the existence of a family of Hopf
categories, with structures closely related to the
quantum groups. In fact, the structure coefficients of the quantum groups
in the canonical bases are positive integers, which can be replaced by
vector spaces. In this paper, we actually construct the Hopf category in
the simplest case, the one corresponding to sl(2), directly.

Our construction of a 4D-TQFT is by no means complete. However, we do have at
least an outline of all the major steps.

The structure of this paper is as follows: In Chapter 1 we recall the
main ideas of the structure of Hopf algebras, and explain how they
translate into 3D-TQFT. Chapter 2 gives the
definition and essential properties of a Hopf category. Chapter 3
contains our new formula, which we call the ``tornado,'' for constructing
a 4D-TQFT from a Hopf category. In Chapter 4 we construct a
non-trivial example of a Hopf category from the quantum
group $U_qsl(2)$ using the canonical bases. Finally in Chapter 5
we discuss mathematical and physical implications.

\smallskip

\newpage

{\bf CHAPTER 1. THE STRUCTURE OF HOPF ALGEBRAS AND 3D TQFT}

\smallskip

The structure of finite dimensional complex Hopf algebras in spite of
its seeming simplicity is rather intricate and has only been gradually
understood during the last three decades.  It turns out that many
known facts and formulas can be elegantly expressed via three
dimensional topological quantum field theory.  We will only recall
those facts and formulas that enter into the topological construction.

Let $A$ be a complex finite-dimensional associative algebra with a
multiplication $M$ and unit $e$, and let $A$ also be an associative
coalgebra (i.e. $A^*$ an associative algebra) with the
comultiplication $\Delta$ and counit $\varepsilon$.  Then $A$ is a
bialgebra if the pairs $(M,e)$ and $(\Delta,\varepsilon)$ satisfy
natural consistency relations:
\begin{align*}
\Delta (xy) &= \Delta (x)\Delta (y), &\Delta e &= e\otimes e ,
\varepsilon (xy) &= \varepsilon (x)\varepsilon (y), &
&\varepsilon (e) &= 1
\end{align*}
where $x,y\in A$.  Clearly, if $A$ is a bialgebra, then $A^*$ is a
bialgebra and vice versa.

Any element $\lambda\in A^*$ defines an associative bilinear form
$(x,y)_\lambda = \lambda (xy), x,y\in A$.  If there exists a
nonsingular $\lambda\in A^*$, i.e. one where the corresponding form is
nondegenerate, then the algebra $A$ is called Frobenius.  Now let $A$
be a bialgebra and let $\lambda\in A^*$ also define a structure of
a Frobenius algebra.  If we require a natural consistency condition
$$
\chi\lambda = \chi (e)\lambda, \quad \chi\in A^*
$$
then we arrive at the notion of a Hopf algebra.  The element $\lambda$
is called a left integral.  In fact, a result of Larson and Sweedler
[18] asserts that the existence of a nonsingular left integral
$\lambda$ in a bialgebra $A$ is equivalent to the existence of the
antipode $s$.  As a corollary, the existence of such a $\lambda$
implies the existence of a nonsingular left integral $\ell\in A$.
Thus a Hopf algebra can be viewed as a Frobenius bialgebra with certain
natural consistency relations.

The above discussion remains true if we replace a left integral
$\lambda\in A^*$ by a right integral $\rho\in A^*$ satisfying
$$
\rho\chi = \chi (e)\rho, \quad \chi\in A^*.
$$
We also denote by $r$ a right integral in $A$.If all four integrals
exist they are unique up to a scalar factor.

Let us define the left and right action of $A$ on $A^*$ by the
formulas
$$
(x\rightharpoonup \chi)(y) = \chi (yx), \quad ( \chi \leftharpoonup x)(y) =
\chi (xy)
$$
where $x,y\in A$ and $\chi\in A^*$.  We have the following
expressions of the antipode and its inverse in terms of the integrals
\begin{align*}
s(x) =\frac{1}{\lambda(\ell)}  (\lambda\leftharpoonup
x)\rightharpoonup\ell &= \frac{1}{\rho (r)} r\leftharpoonup
(x\rightharpoonup\rho) \\
s^{-1}(x) =
\frac{1}{\lambda(r)}
(x\rightharpoonup\lambda)\rightharpoonup r &= \frac{1}{\rho
(e)}\ell\leftharpoonup (\rho\leftharpoonup x)
\end{align*}

Note that the normalization factors are always nonzero numbers.  Next
we introduce a special group-like element $a\in A$, i.e. $\Delta a =
a\otimes a, s(a) = a^{-1}$, by any of the two equivalent conditions
$$
\lambda\chi = \chi (a)\lambda, \quad \chi\rho = \chi (a^{-1})\rho,
\quad \chi\in A^*.
$$
Using the duality between $A$ and $A^*$ we also define a second
group-like element $\alpha\in A^*$.  These two group-like elements
and the number $\alpha (a)$, which turns out to be a root of unity,
play an important structural role.  In particular, one has
\begin{align*}
s(\lambda ) &= a\rightharpoonup\lambda  & s(\rho) &= \rho\leftharpoonup
a^{-1} , \\
s^2(\lambda) &= \alpha (a)\lambda, &s^2(\rho) &= \alpha (a)\rho
\end{align*}

One of the central structural facts about Hopf algebras is the
Radford formula [19]
$$
s^4(x) = a^{-1}(\alpha\rightharpoonup x\leftharpoonup\alpha^{-1})a,
\quad x\in A.
$$
Kuperberg [20] defined an important special subclass of balanced
Hopf algebras for which one has
$$
s^2(x) = a^{-1}x a = \alpha\rightharpoonup
x\leftharpoonup\alpha^{-1}, \quad x\in A.
$$
This subclass is important for the following two reasons:  first
``halves'' of quantum groups are balanced Hopf algebras, second
balanced Hopf algebras admit an elegant topological interpretation.
The relation between these two facts leads to a combinatorial
construction of topological quantum field theories in $3D$, which we
believe will be closely related to CSW theory.

An important property of balanced Hopf algebras is a quasisymmetry of
the forms defined by left and right integrals
$$
\lambda (xy) = \lambda (y(x\leftharpoonup\alpha^{-1})), \quad \rho
(xy) = ((\alpha\rightharpoonup y)x).
$$
For general Hopf algebras, the analogue of the quasisymmetry relation has
a more complicated form.

All the above formulas simplify significantly in the involutive case
$s^2 =id$.  It was shown by Larson and Radford [21] that this
condition is equivalent to the semisimplicity of $A$, and therefore
of $A^*$.

In this case the group-like elements $a$ and $\alpha$ coincide with
units in $A$ and $A^*$, respectively.  Thus left and right integrals
coincide and they can be normalized in such a way that
$$
\lambda^2 = \lambda, \quad \ell^2 = \ell,\quad \lambda(\ell) = (\dim~
A)^{-1}.
$$
As we will explain below, the semisimple case admits an especially simple
topological realization.

\smallskip

{\bf Three Dimensional TQFT and Hopf Algebras}

\smallskip

The labyrinth of identities we have just reviewed, at least
in the case of balanced (and especially semisimple) Hopf algebras, can be
visualized by means of three dimensional topology.  One could also say that the
combinatorics of three dimensional triangulations naturally
leads to the identities for Hopf algebras. This approach to
constructing a 3D-TQFT is originally due to Kuperberg [7,20].
Kuperberg's work is very important, although he did not quite finish the
non-involutory case, since he did
not find a good normalization, and could not control a projective
factor of a root of unity.
We sketch a slightly modified version of it here.

The main idea is to begin with a triangulation of a 3-manifold.
(Actually, any polyhedral decomposition will do as well.)
One labels the edges of the faces (combinatorial 1-2 flags) with
elements of a basis for the algebra, combines around faces with
the multiplication, and around edges with the comultiplication, and
sums over labellings. Then the requirement
of topological invariance will encode the identities of a Hopf algebra.

In order to absorb various powers of group like elements $a$ and
$\alpha$ we define an infinite sequence of integrals by
$$
\mu^{n-\frac{1}{2}} = a^n\rightharpoonup\lambda, \quad n\in\mathbb{Z}.
$$
In fact, since the order of $a$ and $\alpha$ is finite, we get a
cyclic family.  By definition $\mu^{-\frac{1}{2}}$ is a left integral
and $\mu^{\frac{1}{2}}$ is a right integral.

In order to actually define an invariant associated to a triangulated
or polyhedrally decomposed 3-manifold from a balanced Hopf algebra, we need to
associate some
decorations to the decomposition. Specifically, we attach an integer
to each edge, a half-integer and an orientation to each face, and a
preferred vertex (an ``origin'') to each face as well.

\smallskip

Figure 3

\smallskip

{\bf Definition} Let A be a finite dimensional Hopf algebra, B a basis
for it, and let (M, T) be a triangulated 3-manifold. A coloring of (M,
T) by A in the basis B is a function which assigns to each pair of one
face and one incident edge in (M, T) a basis element. we denote the
set of colorings of (M,T) by A as $\Lambda (M,T,A,B)$. We will
frequently suppress the mention of the basis.

Let us denote a coloring of the edges of one face by the elements of
the basis of $A: x_{i_{1}},\ldots ,x_{i_{n}}$.  Then to this polygon P
(as in Figure 3) and to each coloring $\lambda \in \Lambda$ we attach the
following number
$$
\#(P,\lambda)=\mu^r (s^{k_{1}}(x_{i_{1}})\ldots s^{k_{n}}(x_{i_{n}}))
$$

The antiautomorphism property of the antipode guarantees that the
simultaneous shift $k_j\to k_j+1$ together with a change of orientation
and the sign of $r$ changes this number by a factor of $\alpha (a)$.
Clearly repeated application of this shift preserves the orientation
and the sign of $r$.In the case of a semisimple Hopf algebra, the
decoration of the polygon can be simplified: namely, one does not need to
specify a special vertex and a semiintegral number $r$, also the
integers attached to edges are defined $mod ~2$.

Since we are coloring combinations of edges incident to faces, we have
a loop of labels around each edge.  We can think of these as lying on
the edges of a dual polygon, ie a polygon in the dual decomposition to
T. We now form a number for each dual polygon and each labelling. The
formula we use for $\#(E, \lambda)$ is dual to the formula above, i.e. we
use the multiplication and left integral on $A^*$.

To form an invariant of a triangulated $3$-dimensional manifold
$(M,T)$, we form a sum
$$\sum_{\lambda \in \Lambda}\prod_{faces} \# (P, \lambda )\prod_{edges}
\#(E,\lambda),$$
 where we have the following restriction on
decorations. If a polygon intersects a dual polygon then the
difference of the integers attached to the edge of intersection should
be even or odd depending whether the orientations of the polygon and
the dual polygon agree or not.

In the semisimple case, this gives a TQFT. In the more general
balanced case, we need some further restrictions on decorations.
Namely, for any closed contractible contour on the surface obtained as
a regular neighborhood of the 1-skeleton we have a certain linear
relation on the numbers attached to the edges and faces of polygons.
The fact that these restrictions admit solutions follows from the
existence of vector fields on M called combings in [20]. We expect that with a
suitable normalization the
topological invariance of the formula up to a power of $\alpha(a)$
will follow as in [20]. In order to keep track of the power of
$\alpha(a)$, we need further restrictions on decorations given by a
choice of framing rather than a combing.

The invariance of $\# (M,T)$ on the decoration and triangulation
absorbs all the general identities of balanced and semi-simple Hopf
algebras.

In the interest of self-containedness, let us give an outline of the
standard process of constructing a TQFT from a topologically invariant
state sum [8].
First, we assign to any triangulated N-1 dimensional manifold M the
vector space of formal linear combinations of all ways to label the
simplices in the N-1 manifold as triangulated. For any two different
triangulations of M, we pick any triangulation of the N-manifold with
boundary $M\times I$ which agrees with the two triangulations at the
two ends. Next, we do the state sum on $M\times I$ for each choice of
labels on the ends, summing over the interior. This gives us a set of
matrix elements for a map between the two vector spaces associated to
the two triangulations. Identifying under these maps gives us a real
invariant vector space associated to M. It then follows from very
general arguments that we obtain a TQFT.

\smallskip

{\bf CHAPTER 2. THE DEFINITION OF HOPF CATEGORIES}

\smallskip

{\bf Categories and Algebras}

\smallskip

We are now going to construct an analog of  the structure of a Hopf
algebra on a tensor category of a very special type.

To explain the idea of an analog of an algebraic structure on a
category, let us consider first the category {\bf VECT} of finite
dimensional vector spaces.
This category possesses two product operations, $\oplus$ and $\otimes$
and special objects {\bf 0} and {\bf 1} and the following isomorphisms:

$(A\otimes B) \otimes C \Leftrightarrow A\otimes (B\otimes C)$

$(A \oplus B)\oplus C \Leftrightarrow A\oplus (B \oplus C)$

$A\otimes (B \oplus C) \Leftrightarrow (A\otimes B) \oplus (A\otimes
C)$

$A \oplus {\bf O} \Leftrightarrow A$

$A \otimes {\bf 1}\Leftrightarrow A$

$A \oplus B \Leftrightarrow B \oplus A$

These isomorphisms satisfy certain equations, called coherence relations.This
is completely parallel to the definition of a ring.
We describe this by saying that
{\bf VECT} is a ring category. This structure is a
categorical analog of a ring.

There is a natural extension of this category which we call {\bf
Z-VECT}, the category of Z graded vector spaces. Objects in this
category have Z-graded dimensions, which we like to think of as finite
Laurent polynomials in a formal variable. We will also use the category {\bf
Z/n-VECT} of cyclically graded vector spaces.

What has happened is that equations in the ring correspond to
isomorphisms in the category. There are then natural equations
that the isomorphisms should satisfy, so that combining them in
different orders to produce a larger isomorphism always gives
consistent results. These were termed ``coherence relations''
by MacLane [23]. The coherence relations corresponding to the
commutative and associative laws are the Stasheff hexagons and
pentagons [14].

Thus, if we replace an algebraic structure by a categorical analog, we
replace its axioms by a new set of more complex equations, which are
its coherence relations. One of the fundamental ideas which we would
like to emphasize is that if we start with an algebraic structure
which can be used to construct a TQFT, than the coherence relations of
a categorical analog of it are just right to construct a TQFT in one
higher dimension. Our use of a Hopf category in 4D-TQFT is an
application of this idea.

\smallskip

{\bf Hopf Categories.}

\smallskip

Now let us describe the structure of a Hopf category.

\smallskip
{\bf Definition}. A category is semisimple if each object is a direct
sum of simple objects. A semisimple category is finitely generated if it has
only
finitely many inequivalent irreducible objects.

In this paper, we will
have a special interest in finitely generated categories (in order to make all
sums finite).

\smallskip

{\bf Definition}. If {\bf R} is a ring category, then {\bf M} is a
module category over {\bf R} if {\bf M} has an associative direct sum
and we are given a functor ${\bf R \times M
\rightarrow M}$ (denoted as multiplication) such that

\smallskip

$A_1 \otimes ( A_2 \otimes R) \Leftrightarrow (A_1 \otimes A_2)
\otimes R$

\smallskip

$(A_1 \oplus A_2) \otimes R \Leftrightarrow (A_1 \otimes R) \oplus (A_2
\otimes R)$

\smallskip

$A \otimes (R_1 \oplus R_2) \Leftrightarrow (A \otimes R_1) \oplus
(A \otimes R_2)$

and the isomorphisms satisfy the natural coherence relations.
\smallskip

The concept of module category is the categorical analog of the
concept of a module.

\smallskip

{\bf Definition}. An algebra category is a ring category which is also
a {\bf VECT} module such that the structure functors are module maps.

\smallskip

Note that the sets of morphisms between pairs of objects in a {\bf
VECT} module must possess a vector space structure, and that
composition is bilinear.
{\bf VECT} modules are a categorical analog of vector spaces, so
algebra categories are categorical analogs of algebras.

Now recall that the dual of  a category has the same objects as the
category, but with morphisms reversed. Similarly, the dual of any
algebraic construction has diagrams corresponding to the first one,
but with arrows reversed. The dual of a {\bf VECT} module has a
natural
structure as a {\bf VECT} module, including a natural direct sum.

{\bf Definition}. A coalgebra category is a {\bf VECT} module category
whose dual is an algebra category.

\smallskip

This is equivalent to a {\bf VECT} module category with a
comultiplication functor $\Delta :{\bf A}\rightarrow {\bf A\times A}$
satisfying the dual of the axioms of an algebra category
(coassociativity, etc.).

{\bf Definition}. A bialgebra category (without unit and counit) is a {\bf
VECT} module category
which is both an algebra and a coalgebra category, together with a
consistency natural transformation:

\smallskip

Figure 4

\smallskip

The associativity and coassociativity isomorphisms and the consistency

map $\alpha : \Delta(A) \otimes \Delta(B) \Leftrightarrow \Delta
(A\otimes B)$ must satisfy the following four commuting cubes as
coherence relations:

\smallskip

Figure 5

\smallskip

The first two of these are the Stasheff pentagon and the dual object
for comultiplication. The latter two are new.

Finally, a  Hopf category is a semisimple bialgebra category together with the
categorical analogs of a unit, counit, Frobenius inner product and
dual inner product. For brevity, we omit discussion of the
corresponding coherence relations.

For each of the categories we have discussed above, we can form its
Grothendieck ring by using the direct sum and tensor product to define
additions and multiplications of formal linear combinations of objects
in the category. One then recovers an algebra with structure analogous
to the category.
\smallskip

{\bf The Structure Constants of A Hopf Category.}

\smallskip

The above definition is extremely abstract, and hence not that
convenient to use in writing a numerical formula. We want to think of the
associativity,
coassociativity, and consistency isomorphisms of a Hopf category as
determining some numbers as ``matrix elements,'' which must satisfy
certain axioms as a result of the four cubes. It will then transpire
that these numbers can be naturally fitted into a triangulated
4-manifold, and that the properties they satisfy guarantee topological
independence.

The idea of extracting a number from the associator of a tensor
category is one which is familiar to physicists and chemists under the
guise of 6J symbols [16]. The classical setting for this idea is the
representation category of the Lie group SU(2), which is used to
understand the addition of spins in quantum mechanics.

The associator of a tensor category gives us a map
\smallskip

$A\otimes(B \otimes C) \Rightarrow (A \otimes B) \otimes C$.

\smallskip

In order to extract a minimal unit of information from this in a
semisimple category, we let A, B, and C be irreducible objects,
denoted $ R_1, R_2, R_3$. We then pick irreducible subobjects $R_{12}
\subset R_1 \otimes R_2, R_{23} \subset R_2 \otimes R_3; R_{123}
\subset R_1 \otimes R_2 \otimes R_3$. These imply a choice of four
tensor operators $ \mu : R_1 \otimes R_2 \rightarrow R_{12},
\nu :R_{12} \otimes R_3 \rightarrow R_{123}; \rho: R_2 \otimes R_3
\rightarrow R_{23}$, and $ \sigma: R_1 \otimes R_{23} \rightarrow
R_{123}$.

In order to define the 6J symbols in the simplest (although not the
most invariant) way, let us assume that we have chosen bases for the
space of tensor operators for each triple of simple objects in our
category, and that our four choices are basis elements.

Now we can form $\nu \circ \mu$ and $\sigma \circ \rho$. These are
elements of two different bases for the same space, namely the tensor
operators from $R_1 \otimes R_2 \otimes R_3$ to $R_{123}$. The coefficient of
the first in the expansion of the second in the first basis is a scalar, and is
defined as a 6J symbol.

We can graphically represent the 6J symbol as follows:

\smallskip

Figure 6

\smallskip

The coherence cube implies that the associator can be combined
to form more complex quantities called 3NJ symbols. There is a 3NJ
symbol for every labelling of the triangulated boundary of a
polyhedron, where edges are labeled with objects in the category and
faces are labelled with tensor operators. The edges are directed, and
the tensor operators, strictly speaking, depend on the directions, but
can be reversed in a natural way using the duality of the category.
Thus a triangle labelled

\smallskip

Figure 7

\smallskip

represents a tensor operator $R_1 \otimes R_2 \otimes R_3 \rightarrow
{\bf 1}$; while a triangle

\smallskip

Figure 8

\smallskip

represents an operator $R_1 \otimes R_2 \rightarrow {R_3}^*$. These
can be naturally identified, however.

A 3NJ symbol can be evaluated by cutting up the interior of the
polyhedron into tetrahedra, labelling internal edges with objects and
internal faces with tensor operators, taking the product of the
resulting 6J symbols on the tetrahedra, then summing over the internal
labellings. The result, by an application of the coherence cube, is
independent of the choice of triangulation. This is a well known
technique in the chemical literature for the case of SU(2), where it
is called the graphical calculus [16].

Similarly, we have dual $6J*$ and $3NJ*$ symbols corresponding to
the coproduct and its coassociator.

The quantities produced by the graphical calculus are closely related
to the construction of a 3D-TQFT by Turaev and Viro [8]. The
construction first described by those two authors really does not
require a braided tensor category, only a tensor category with a very
weak condition on duals. One sufficient condition, perhaps not the
weakest, is called ``spherical'' in [24].

Finally, the consistency isomorphism of a bialgebra category gives us
a new set of numbers, which we call $3\times3J$ symbols.

In order to specify a $3\times3J$ symbol, we pick nine irreducible
objects in the (finitely generated, semisimple) category $R_{ij}$,
i,j=1,2,3. In addition, we pick ``tensor operators'' $\tau_i :
R_{i1}\otimes R_{i2} \rightarrow  R_{i3}$ and ``cotensor operators''
$\kappa_j :R_{1j}\Box R_{2j} \rightarrow \Delta(R_{3j})$
, where we use the symbol $\Box $ to denote an object in $A\times A$,
rather than the tensor product of our bialgebra category A.

This gives us a number, because we can think of our data as giving us
one map from $\Delta(R_{11} \otimes R_{22} )$ to $R_{33}$ and another
from $\Delta R_{11} \otimes \Delta R_{22}$ to the same target. The
coefficient of one in the expansion of the other in the first basis
after composing with the compatibility isomorphism gives
us the $3\times 3 J$ symbol. In order to clarify this, let us introduce a
graphical calculus, similar to the 3D-TQFT in
[25].

Let us draw words in our bialgebra category as two dimensional
triangular complexes. Let us place objects of our category on edges of
faces (combinatorial 1-2 flags), and represent the comultiplication as
a hinge where three triangles meet, and the multiplication as two
edges of a triangle combining to produce a third.

In this language, we can draw the consistency isomorphism as follows;

\smallskip

Figure 9

\smallskip

This is closely related to an equation in the 3D-TQFT in [25]. Now on
our
new categorical level, it is a morphism, and should really be thought
of as a 3D complex in 4 dimensions, a movie of a cup being squashed.

In this picture, the 3 tensor operators live on the 3 triangles,
while the 3 cotensor operators inhabit the 3 hinges. This
provides a topological picture for $3\times3J$ symbols.

There is a natural family of combinations of $3\times3J$ symbols,
which we call $N-3\times3J$ symbols. To specify one of these, we need
3N irreducible objects in the category, $R_{ij}$ i=1,2,3; j=1,2...N;
together with cotensor operators $\tau_j: R_{1j}\otimes
R_{2j}\rightarrow R_{3j}$ and $\kappa_i : R_{i2}\Box
R_{i3}...R_{iN}
\rightarrow \Delta^N(R_{i1})$.

We draw these as follows:

\smallskip

Figure 10

\smallskip

They are defined by breaking up the polygons into triangles, and
summing over all labellings of the internal lines and triangles with
triplets of irreducible objects joined by tensor operators and
triplets of cotensor operators respectively, multiplying the resulting
$3\times3J$ symbols, and adding over labellings where the products of
the cotensor operators on the triangles give the $\kappa_i$.

The coherence cube then assures us that the result is independent of
the choice of triangulation of the polygon.

There is also a dual family of $N-3\times3J*$ symbols, reversing the
roles of tensor and cotensor operators in the above construction.

We refer to this property as the thickened associativity of
$3\times3J$ symbols.

\smallskip

{\bf CHAPTER 3. THE TORNADO FORMULA}

\smallskip

The structural elements of a Hopf category we described above have a
natural relationship with the topology of triangulated 4-dimensional
manifolds.

{\bf Definition}. Let M be a 4-manifold, and T be a triangulation of M.
Furthermore, let A be a finitely generated, semisimple Hopf category.
A labelling of (M,T) with respect to A is an assignment of an
irreducible
object of A to each pair of an edge incident to a tetrahedron,
together with an assignment of a
basis element of the space of tensor operators to each pair of a face incident
to a tetrahedron,
joining the objects assigned to the edges of the triangle and the
tetrahedron; and an assignment of a basis vector of the space of cotensor
operators to each pair of
an edge and a face around the loop of objects assigned to the loop of
of tetrahedra containing the triangle.

Strictly speaking, the above definition is slightly ambiguous, since
we need to specify which objects are in the domain and range of
the tensor or cotensor operator. The dualities of the category allow
us to identify different choices of this, for example $ Hom( A\otimes
B, C) \Leftrightarrow Hom (A, B^*\otimes C)$. We really need to orient
our manifold and order the vertices, then use a convention to remove
this ambiguity. It will then be necessary to prove that our invariant
formula is independent of the choice of ordering. In the interest of brevity of
notation and exposition,
we will omit this wherever possible.

Let us denote by $\bigwedge (A,M,T)$ the (finite) set of labellings of
(M,T) with respect to A.

Now let us note that each $\lambda \in \bigwedge(A,M,T)$ assigns to
each
tetrahedron t in the triangulation a 6J symbol $6J(t,\lambda)$, to
each
triangle f a $N-3\times3J $ symbol $ N-3\times 3J(f, \lambda)$, and to
each edge e a $3NJ*$ symbol $3NJ*(e,\lambda)$.

In this interlocking set of assignments, each tensor and cotensor operator
appears in two places, once
corresponding to a triangle, and once on an edge or tetrahedron.

We can now form the following expression:

$TORN(A,M,T)=$

$\sum_{\lambda \in \bigwedge(A,M,T,)} \prod_e
3NJ*(e,\lambda)\prod_f N-3\times3J(f.\lambda)\prod _t6J(t,\lambda)$

This expression, which we are calling the tornado, (partly because the
configuration of labels around a triangle in 4-space reminds us of a
vortex,), is a completely natural
analog of the 3D formula. The sum seems to depend on our
choices of bases for our spaces of tensor and cotensor operators.
However,  each operator actually appears twice, and if we include a
convention using orientation as to exactly how it appears, the two
appearances are in dual spaces. Thus, the
dependence on the choice of basis actually cancels itself out.

\smallskip

{\bf THEOREM 1}. The tornado formula is independent of the
triangulation, and hence gives an invariant of 4-manifolds. If we take
a relative version of the tornado formula on manifolds with boundary,
it gives us a 4D-TQFT.

\smallskip

{\bf CONJECTURE 1}. The tornado formula admits an extension to the case
of a non-involutory Hopf category.

\smallskip

We believe that the extension will be analogous to the 3D case we
discussed in Chapter 1.

In this paper we will give the outline of a proof of theorem 1. A more
complete treatment will appear in a later paper. We briefly discuss
the role of conjecture 1 in the conclusions.

\smallskip

{\bf Outline Of a Proof of Topological Invariance.}
\smallskip

Let us first outline a general strategy to prove that sums like the
tornado are invariant, then show how the properties of a Hopf category
allow us to carry out this method for the tornado.

A summation of the type of the tornado formula, where labels are
attached someplace in a triangulation (for example, to combinatorial
flags, or to simplices of certain dimensions), and some sort of combinations
are computed (on other dimensional simplices), after which the labelling is
summed over, is called a
state sum.

{\bf Definition}. We say a state sum has the blob property if the sum,
when carried out on any triangulation of a ball, with the values of
labels in the boundary of the ball fixed and the labels in the
interior of the ball summed over, equals the value of the
contributions from the boundary of the ball only.

The blob property can be thought of as a sort of Poincare lemma.

{\bf Lemma}. A state sum with the blob property gives an invariant of
closed manifolds. If applied to manifolds with boundary, it gives rise
to a TQFT.

Outline of proof: Any two triangulations of an n dimensional manifold
can be related by a series of elementary moves called ``bistellar
equivalences'' due to Pachner [26]. The moves all consist in replacing
one configuration with another with the same boundary. Thus, the blob
lemma implies invariance under Pachner's moves. This means that a
state sum with the blob property gives a value on a closed manifold
which does not depend on the triangulation. There is a standard method
for extending this to a TQFT by doing the state sum on triangulated
manifolds with boundary [8], as we briefly explained in chapter 1.

Now let us explain a general strategy for showing that a certain type
of state sum has the blob property. By induction, it suffices to show
that the join along some part of its boundary of a single simplex to a
ball with some triangulation, if summed only over boundary
contributions of the ball, and all contributions from the simplex,
gives the same result as a sum over the boundary of the new larger
ball resulting from the union.

\smallskip

Figure 11

\smallskip

In order to prove this equivalence, we can break it up into a series
of two parts.

1.Rearrangement. We need to change the triangulation of the shared
part of the boundary so it becomes identical to the other half of the
boundary of the simplex.

2.Squashing. We need to be able to collapse down the ``pillow'' between
one pair of overlapping simplices at a time.

The picture of the consistency isomorphism we drew above is an example
of a squashing move. In the proof of the topological invariance of the
3D-TQFT constructed from a Hopf algebra in [25], this move plays
essentially this role.

Now let us consider how to apply this strategy to the tornado formula.
We need to be able to carry out two kinds of moves on our summation,
rearrangements and squashings. Basically, the rearrangements follow
from the coherence cubes of the multiplication; while the squashing
move is  one version of the coherency cube of associativity with the
consistency isomorphism.

\smallskip

Figure 12

\smallskip

As in the three dimensional case, we expect that the invariance of the
tornado will use all the identities of Hopf categories corresponding
to semisimple Hopf algebras. Assuming conjecture 1, a similar relation
should hold for balanced Hopf algebras as well.

{\bf Chapter 4. THE HOPF CATEGORY ASSOCIATED TO THE HOPF ALGEBRA
$u^+_q(\frak{sl}_2)$ }

\smallskip

It is well-known that the quantum groups $U_q(\frak g)$ gives rise to
a rich class of $3D$-TQFTs, which can be viewed as  combinatorial
counterparts of the CSW path integrals.  To construct a $3D$-TQFT directly
from a  Hopf algebra, one only needs ``half'' of the quantum group, ie
$U^+_q(\frak g)$.  This approach presumably leads to the same CSW
theory and avoids the preparatory work of constructing a modular tensor
category.  When $\frak g = \frak{sl}_2$, the Hopf algebra
$U^+_q(\frak g)$ over $\bold C[q,q^{-1}]$ is generated by $e, q^{\pm
h}$ with the relations
$$
q^h q^{-h} = 1, \quad q^h e = eq^{h+2}.
$$
The comultiplication is given by
$$
\Delta q^h = q^h\otimes q^h, \quad \Delta e = e\otimes 1 + q^h\otimes
e.
$$
In order to obtain a finite-dimensional Hopf algebra we have to
consider the case when $q$ is a root of unity.  We set $q = exp(2\pi
i/2r)$ and we impose extra relations
$$
e^r = 0, \quad (q^h)^{2r} = 1.
$$
As a result, we obtain a finite dimensional Hopf algebra, which will
be denoted $u^+_q(\frak{sl}_2)$.

Before we proceed to the construction of a Hopf category associated to
$u^+_q(\frak{sl}_2)$ we will first discuss integral bases.  For
example, one can choose a basis $q^{nh} e^m, n\in\bold Z, m\in\bold
Z_+$.  Then the structure coefficients of multiplication and
comultiplication belong to $\bold Z_+[q,q^{-1}]$.  This basis would
be sufficient for the purpose of this paper.  However, we will prefer
to consider another basis, which can also be extended to the full
quantum group $U_q(\frak{sl}_2)$.

Following Lusztig [27] we will introduce a modified Hopf algebra
$\dot U^+_q(\frak{sl}_2)$ with generators $e_n, 1_n, n\in\bold Z$,
satisfying the relations
$$
1_m1_n = \delta_{m,n}1_m, ~~e_m1_n = \delta_{m-1,n}e_m, ~~ 1_ne_m =
\delta_{m+1,n} e_m.
$$
Then if we define the comultiplication by
$$
\Delta 1_n =\sum\limits_{n'+n''=n} 1_{n'}\otimes 1_{n''}, ~\Delta e_n
= \sum\limits_{n'+n''=n} (e_{n'}\otimes 1_{n''} + q^{n'}
1_{n'}\otimes e_{n''})
$$
we obtain a Hopf algebra denoted $\dot U^+_q(\frak{sl}_2)$.  Setting
$$
e = \sum\limits_{n\in\bold Z} e_n, \quad q^{\pm h} =
\sum\limits_{n\in\bold Z} q^{\pm n} 1_n
$$
in an appropriate completion of $\dot U^+_q(\frak{sl}_2)$ one
recovers the Hopf algebra $U^+_q(\frak{sl}_2)$.  We will consider the
following basis in $\dot U^+_q(\frak{sl}_2)$
$$
e^{(m)}_m = \frac{1}{[m]!} e_{n+m-1}e_{n+m-3}\ldots e_{n-m+1}, \quad
n\in\bold Z, m\in\bold Z_+
$$
where $[m]! = [1][2]\ldots [m], [m] = (q^m-q^{-m})/(q-q^{-1})$.  One
has $e^{(0)}_n = 1_n, e^{(1)}_n = e_n$.  Then the multiplication and
comultiplication in this basis admits the following form:

$$
e_n^{(m)}e_{n'}^{(m')}=\left[\begin{array}{c} m+m'\\
m\end{array}\right]\delta_{n-m',n'+m}e_{n-m'}^{m+m'}\leqno(1)
$$
\smallskip

$$
\triangle
e_n^{(m)}=\sum_{\stackrel{m'+m''=m}{n'+n''=n}}q^{n'm''}e_{n'}^{(m')}\otimes
e_{n''}^{(m'')}\leqno(2)
$$

\bigskip

Thus the structure coefficients of multiplication and comultiplication
belong to $\bold Z_+[q,q^{-1}]$.  Associativity of multiplication and
comultiplication and their consistency yield, respectively the
following identities in $\bold Z_+[q,q^{-1}]$

$$
\left[\begin{array}{c}
m+m'\\ m\end{array}\right]
\left[\begin{array}{c}
m+m'+m''\\ m+m'\end{array}\right]=
\left[\begin{array}{c}
m+m'+m''\\ m\end{array}\right]
\left[\begin{array}{c}
m'+m''\\ m'\end{array}\right]\leqno(3)
$$
\vskip .10in
$$
q^{(n'+n'')m'''}q^{n'm''}=q^{n'(m''+m''')}q^{n''m'''}\leqno(4)
$$
\vskip .10in
$$
\left[\begin{array}{c}
m+m'\\ m\end{array}\right] =\sum_{m_{1}+m_{2}=m}q^{\tilde
mm_{2}-\tilde m'm_{1}}\left[\begin{array}{c}
\tilde m\\ m_1\end{array}\right]\left[\begin{array}{c}
\tilde m'\\ m_2\end{array}\right]$$, where $$ m+m'=\tilde m+\tilde m'\leqno(5)
$$

Again when $q = exp(2\pi i/2r)$ we can consider a finite dimensional
Hopf algebra $\dot u^+_q(\frak{sl}_2)$ with the basis $e^{(m)}_n,
m=0,1,\ldots ,r-1, n\in\bold Z/2r\bold Z$.

\smallskip

{\bf The combinatorial construction of the Hopf category. }

\smallskip

To construct a Hopf category corresponding to $\dot
U^+_q(\frak{sl}_2)$ we replace the structure coefficients by graded
vector spaces of the corresponding dimension.   Let $V =
\bigoplus\limits_{k\in\bold Z} V_k$ be a finite dimensional graded
space.  We define its graded dimension by
$$
\dim_q V = \sum\limits_{k\in\bold Z} q^k \dim ~V_k.
$$
We also define a shift operator $T$ that shifts degree of elements by
$1$ and its inverse $T^{-1}$.  Clearly one had $\dim_q T^nV = q^n
\dim_qV, n\in\bold Z$.  For each $n\in\bold Z$ we define a quantum
$n$-dimensional vector space $V^n$ with a basis $\{v_k\}$ indexed by
degree $k\in\{n-1,n-3,\ldots, -n+1\}$.   We will identify
$V^1\cong\bold C$.  Then we have $\dim_q V^n = [n]$.  Finally  we
denote by $\Lambda^mV^n$ the $m$-th exterior power of $V^n$, it has a
natural basis $v_{k_{i}}\wedge\ldots\wedge v_{k_{m}} - n < k_1
<\ldots < k_m < n$.  One can easily show that $$\dim_q \Lambda^mV^n =
\left[\begin{array}{c} n\\m\end{array}\right]$$.

A complimentary choice of indices establishes an isomorphism
$\Lambda^mV^n\simeq \Lambda^{n-m}V^n$. Now we are ready to replace the
identities for the structure constants by the following linear isomorphisms
$$
\alpha:\Lambda^m
V^{m+m'}\otimes\Lambda^{m+m'}V^{m+m'+m''}\longrightarrow
\Lambda^mV^{m+m'+m''}\otimes\Lambda^{m'} V^{m'+m''} \\
$$
and
$$
 \beta: T^{n'm''}\bold C\otimes T^{(n'+n'')m'''}\bold
C\longrightarrow  T^{n'(m''+m''')}\bold C\otimes
T^{n''m'''}\bold C \\
$$
and
$$
\gamma: \Lambda^m
V^{m+m'}\longrightarrow\bigoplus\limits_{m_{1}+m_{2}=m}
T^{\tilde m m_{2}-\tilde m'm_{1}}
\Lambda^{m_{1}}V^{\tilde m}\otimes \Lambda^{m_{2}} V^{\tilde m'},
m+m' = \tilde m+\tilde m'
$$
where $\alpha,\beta, \gamma$ admit a simple combinatorial
description.  One can break the basis of $V^{m+m'+m''}$ into three
subsets of order $m, m'$ and $m''$, respectively, either by choosing
first a subset of order $m+m'$ and inside it a subset of order $m$ or
by choosing first a subset of order $m'+m''$ and inside it a subset
of order $m'$.  The comparison between these two ways yields the
isomorphism $\alpha$.  The isomorphism $\beta$ is self-evident.  To
construct the isomorphism $\gamma$, we first break the basis of
$V^{m+m'}$ into two segments of order $\tilde m$ and $\tilde m'$,
correspondingly.  Then any choice of $m$ basic elements in $V^{m+m'}$
is automatically partitioned into two parts (say of orders $m_1$ and
$m_2$) corresponding to the two segments. Since the degrees of
elements in both segments are shifted in comparison with standard
degrees of $V^{\tilde m}$ and $V^{\tilde m'}$, we need an adjustment
given by $T^{\tilde m m_{2}-\tilde m' m_{1}}$ .

We define a semisimple $\bold Z$-VECT module category $\dot{\frak
U}_q(\frak{sl}_2)$ by indicating a set of simple objects $E^{(m)}_n,
m\in\bold Z_+, n\in\bold Z$, with $Hom(E^{(m)}_n,
E^{(m')}_{n'})=\bold C$ if $m=m', n=n'$ and zero otherwise.  We also
define monoidal $\boxtimes$ and comonoidal $\triangle$ structure in
$\dot{\frak U}_q^+(\frak{sl}_2)$ by
$$
E_n^{(m)}\Box
E_{n'}^{(m')}\stackrel{\sim}{\Rightarrow}\left\{\begin{array}{l}
\wedge^m\vee^{m+m'}\otimes E_{n+n'}^{m+m'},n-m'=n'+n\\
0~,~{\rm otherwise}\end{array}\right.\leqno(6)
$$
$$
\triangle
E_n^{(m)}\stackrel{\sim}{\Rightarrow}\bigoplus\limits_{m'+m''=m}T^{n'm''}{\bf
C}\otimes E_{n'}^{(m')}\Box
E_{n''}^{(m'')}\leqno(7)
$$
Then the linear isomorphism $\alpha$ induces an associativity
isomorphism, $\beta$ induces a coassociativity isomorphism and $\gamma$
provides $\dot{\frak U}^+_q(\frak{sl}_2)$ with the structure of
a bialgebra category.  We obtain

{\bf Theorem 2}.  The semisimple $\bold Z$-VECT module category
$\dot{\frak U}^+_q(\frak{sl}_2)$ provided with monoidal and comonoidal
structures as above as well as associativity, coassociativity and
consistency isomorphisms is a bialgebra category.

{\bf Proof}.  One needs to verify the pentagon relations for
the associativity and coassociativity isomorphisms and their relations
with the consistency isomorphism.  This follows from the
combinatorial descriptions of linear maps $\alpha, \beta, \gamma$.
In fact, the pentagon for associativity follows from an analysis of the
subdivision of the basis of $V^{m+m'+m''+m'''}$ into subsets of order
$m, m', m''$ and $m'''$ and the definition of $\alpha$.  The relation between
$\alpha$ and $\gamma$ follows from an analysis of a subdivision of the basis
of $V^{m+m'+m''}$ into parts of order $m,m'$ and $m''$ relative to
segments of order $\tilde m$, and $\tilde m'$.  Similarly, the relation
between $\beta$ and $\gamma$ follows from an analysis of a subdivision
of the basis of $V^{m+m'}$ into parts of order $m$ and $m'$ relative
to segments of order $\tilde m, \tilde m'$ and $\tilde m''$.  Finally
the pentagon for coassociativity follows from the relation between shifts of
segments of $\tilde m, \tilde m', \tilde m''$ and $\tilde m'''$ in
the basis of $V^m$.

When we pass to the root of unity case, we need to replace the
category of graded vector spaces $\bold Z$-VECT by a category
of $\bold Z/2r$-cyclicly graded vector spaces, then define a
subquotient finitely generated category by a procedure reminiscent of the
factorization of the subcategory of representations of a quantum group by
the ideal subcategory of representations with zero quantum dimension.  In
addition, one needs to introduce categorical analogues to the bilinear forms
in $u^+_q(\frak{sl}_2)$.

One of our goals is the following:

{\bf Conjecture 2:}.  There exists a subquotient of the Hopf
category associated to $U^+_q(\frak{sl}(2))$ with the Grothendieck ring given
by $\dot
u^+_q(\frak{sl}_2)$.

Clearly the combination of conjectures 1 and 2 should provide
a combinatorial construction of four dimensional invariants for any
$r\in\bold Z_+$.  If true, they will yield a natural generalization
of the celebrated $3D$ invariants from quantum groups at roots of unity.

{\bf CHAPTER 5. CONCLUSIONS}

In this paper we have outlined a new approach to constructing 4D-TQFT.
Much work remains before we can say whether this construction can
distinguish smooth structures, or has any relationship to
Donaldson-Floer theory. We expect that in order to find examples of
the tornado formula which are sensitive to smooth structure, it will
be necessary to modify it to include cases of Hopf categories
corresponding to nonsemisimple Hopf algebras, in particular the
quantum groups.

Just as state sums associated to balanced Hopf categories are sensitive
to framings of 3-manifolds, we expect the modified tornado formula to
be sensitive to the choice of a bundle over the 4-manifold. Note that
nontrivial bundles on 4-manifolds can be described via
changes of framings over 3-manifolds which cut them. This would be
similar to the dependence of DF theory on instanton number, which
plays a crucial role in detecting smooth structures.

By analogy with the 3D case, we expect the relevant Hopf categories to
contain grouplike objects, which can be used to insert instantons in
the topological picture.

We synthesize our conjectures 1 and 2 into the following grand conjecture:

{\bf Main Conjecture}. The TQFT arising from the modification of the
tornado formula appropriate to noninvolutive Hopf categories, when
applied to the
Hopf categories associated to quantum groups (described in chapter 4)
is sensitive to smooth structure, and is related to Donaldson-Floer theory.

The combination of ideas we described in the introduction suggest more
than one route towards construction of a 4D-TQFT. On the one hand, we
could form the 2-category of module categories over our Hopf category
[15].
This would form a very special tensor 2-category, which we could use
to construct a 4D theory analogously to the 3D construction in [7].

On the other hand, we believe that our Hopf category itself appears as
a category of representations of a more intricate algebraic structure,
which we call a trialgebra.
A trialgebra should have two multiplications and one comultiplication
satisfying natural consistency relations.

The most interesting examples of trialgebras should have the Hopf
categories associated to the quantum groups for categories of
representations. There is strong evidence that the simple example of a
Hopf category corresponding to the quantum group $U_q(sl(2))$ can be
obtained from certain representations of the Virasoro algebra at c=1,
which are related to the c=1 critical string.

The various algebraic constructions of low dimensional TQFTs fit
together into a picture which we call the dimensional ladder.

\smallskip

Figure 13

\smallskip

A general programmatic discussion of the dimensional ladder will
appear in [28].

\newpage

FIGURE CAPTIONS

\smallskip

1. 2D topological move

2. The associative law

3. A decorated face

4. The consistency isomorphism

5. The coherence cubes of a Hopf category

6. A 6J symbol

7. An oriented triangle

8. A differently oriented triangle

9. Graphic expression of the consistency isomorphism

10. An $ N-3 \times 3J$ symbol

11. Inductive step for blob property

12. 4D squashing move=coherence cube

13. The dimensional ladder

\newpage

{\bf BIBLIOGRAPHY}

\smallskip

[1]  M.F. Atiyah
, Topological quantum field theories
, Publ. Math. Inst. Hautes Etudes Sci., Paris,
vol 68, 1989, p175-186

[2]  E. Witten
, Topological quantum field theory
, Comm. Math. Phys.,
vol 117 , 1988
p353--386

[3] G. Segal, Conformal Field Theory, in Proceedings of the
International conference on Mathematical Physics, Swansea, 1988

[4]  N. Reshetikhin and V.G. Turaev
, Invariants of 3-manifold via link polynomials and quantum
groups
, Invent. math.,
vol 103, p547-597,  1991

[5] L. Crane, 2-d Physics and 3-d Topology, Commun. Math. Phys.,
vol135, p615-640, (1991)

[6] K. Walker, On Wittens 3 manifold invariants, unpublished

[7] G. Kuperberg
, Involutory Hopf algebras and three-manifold invariants,
 Int. J. Math. vol2 p41-66, (1990)

[8]  V. Turaev and O. Viro
, State sum invariants of $3$-manifolds and quantum $6j$ symbols
, Topology
vol 31, p865-902,  1992

[9] M. Fukuma, S. Hosono, H. Kawai
, Lattice topological field theory in two dimensions,
Commun. Math. Phys. vol166, p157-175, (1994)

[10] A. Floer, An instanton invariant for 3-manifolds, Commun. Math. Phys.
vol118,
p215-240, 1988

[11] R. Dijkgraaf and E. Witten, Topological Gauge Theories and Group
Cohomology, Commun. Math. Phys vol129, p393-429, 1990

[12] D. Yetter, Topological quantum field theories associated to finite
groups and crossed G-sets, Journal of Knot Theory and its
Ramifications volI, p1-20, 1992

[13] D. Freed and F. Quinn Chern Simons theory with finite gauge
group,
Commun. Math. Phys. to appear

[14] L. Crane and D. Yetter A Categorical Construction of 4D
Topological Quantum Field Theories, in Quantum Topology, World
Scientific 1993, L. Kauffman ed.

[15] M. Kapranov and V. Voevodsky, Braided Monoidal 2-Categories,
2-Vector Spaces and Zamolodchikov's Tetrahedra Equation,
preprint

[16] A. P. Yutsis, I. B. Levinson  and V. V. Vanagas, Mathematical
Apparatus of the Theory of Angular Momentum (Israel Program for
Scientific Translations, Jerusalem).

[17] G. Lusztig
, Canonical bases arising from quantized enveloping algebras I
, J. Amer. Math. Soc.
vol 3, p447-498, 1990

[18]  R.G. Larson and M.E. Sweedler
, An associative orthogonal bilinear form for Hopf algebras
, Amer. J. Math. vol 91, p75-94, 1969

[19] D.E. Radford
, The order of the antipode of a finite dimensional Hopf algebra
is finite
, Amer. J. Math vol 98, p333-355, 1973

[20] G. Kuperberg A Definition of $\# (M,H)$ in the Non involutory
Case, unpublished.

[21]  R.G. Larson and D.E. Radford pages 187--195
, Semisimple cosemisimple Hopf algebras
, Amer. J. Math vol 109 , 1987

[22] E. Witten, Quantum Field Theory and the Jones Polynomial, Commun.
Math. Phys. 121 p351-399, 1989

[23]  S. MacLane
Categories for the working mathematician,
Springer Verlag
, 1988

[24] J. W. Barrett and B. W. Westbury, Spherical Categories,
University of Nottingham preprint, 1994

[25] S. Chung, M. Fukama, and A. Shapere,
, Structure of Topological Field Theories in Three Dimensions
Int. J. Mod. Phys. A, p1305-1360, (1994)

[26]  U. Pachner
, P.L. Homeomorphic Manifolds are Equivalent by Elementary Shelling
, Europ. J. Combinatorics
vol 12, p129-145,  1991

[27]  G. Lusztig
, Introduction to Quantum Groups
, Birkhauser ,  1993

[28] L. Crane and I.B. Frenkel, A Representation Theoretic Approach to TQFT, to
 appear
in the proceedings of the Gelfand seminar

\end{document}